\documentstyle[11pt,emulateapj,epsfig,psfig,rotate]{article} 

\newcommand{\mdot}{\dot M}

\newcommand{\etal}{{\it et al. \/}}
\newcommand{\be}{\begin{equation} }
\newcommand{\ee}{\end{equation} }
\newcommand{\bea}{\begin{eqnarray} }
\newcommand{\eea}{\end{eqnarray} }
\newcommand{\MS}{M_\odot}
\newcommand{\kes}{\kappa_{es}}
\newcommand{\tes}{\tau_{es}}
\newcommand{\kff}{\kappa_{ff}}
\def\kms{${\rm km\, s}^{-1}$}
\def\kmss{${\rm km\, s}^{-1}$ }

\newenvironment{mylist}[1]{%
\begin{list}{}
    {
        
        \settowidth{\labelwidth}{#1}
        \setlength{\leftmargin}{1.2\labelwidth}
    }
}{%
\end{list}}

\lefthead{Treves \etal}
\righthead{Isolated Neutron Stars}

\begin{document}

\title{Isolated Neutron Stars: Accretors and Coolers}
\author{Aldo Treves}
\affil{Dipartimento di Scienze, Universit\`a  dell'Insubria, \\
Via Lucini 3, 22100, Como, Italy \\ e--mail:
treves@uni.mi.astro.it}
\author{Roberto Turolla}
\affil{Dipartimento di Fisica, Universit\`a di Padova, \\ Via
Marzolo 8,
35131 Padova, Italy \\ e--mail: turolla@pd.infn.it}
\author{Silvia Zane}
\affil{Nuclear and Astrophysics Laboratory, University of Oxford,
\\ Keble Road, Oxford OX1 3RH, England\\
e--mail: zane@astro.ox.ac.uk}
\and
\author{Monica Colpi}
\affil{Dipartimento di Fisica, Universit\`a degli Studi di Milano Bicocca,
\\ Piazza della
Scienza 3,
20126, Milano, Italy \\ e--mail: colpi@uni.mi.astro.it}

\begin{abstract}

As many as $10^9$ neutron stars populate the Galaxy, but only
$\approx 10^3$ are directly observed as pulsars or as accreting
sources in X--ray binaries. In principle also the accretion of the
interstellar medium may make isolated neutron stars shine, and
their weak luminosity could be detected in soft X--rays.
Recent ROSAT observations have convincingly shown that neutron
stars accreting from the interstellar medium are extremely rare, if
observed at all, in contrast with earlier theoretical predictions.
Until now two possible explanations for their elusiveness have
been proposed: their velocity distribution may peak at $\sim 200-400
\ {\rm km\, s}^{-1}$, as inferred from pulsar statistics, and this
would severely choke accretion; the magnetic field may decay on
timescales $\sim 10^8-10^9$ yr, preventing a large fraction of
neutron stars from entering the accretor stage. The search for
accreting neutron stars has produced up to now a handful of promising
candidates. While little doubt is left that these objects are
indeed isolated neutron stars, the nature of their emission is
still controversial. In particular accreting objects can be
confused with much younger, cooling  neutron stars. However, a
combination of observations and theoretical modeling may help in
discriminating between the two classes.

\end{abstract}

\keywords{Stars: Neutron --- Stars: X--rays --- Stars: Dynamics}

\section{Introduction}

Neutron stars (NSs hereafter) are a well established end point of
stellar evolution. Although their existence was proposed in the
thirties, their actual discovery coincides with that of pulsars
and with the recognition that pulsars are young, rapidly spinning
NSs, endowed with a large magnetic field. A few years later it was
realized that most X--ray binaries, and X--ray pulsators in
particular, do contain accreting NSs.

Altogether observed pulsars and X--ray binaries account for $\sim 1500$
objects. However, as indicated by various arguments (see section
4), NSs should represent a non--negligible fraction, $\sim 1\%$,
of the stars in the Galaxy,  with a total number as high as
$10^9$. NSs are therefore, rather numerous but extremely elusive.

Let us consider isolated NSs, which supposedly represent the bulk
of the population. Soon after their birth they may show up as
radio pulsars, if the beaming conditions are favorable, and the
pulsar phase may last some $10^7$ yr. At formation NSs are hot ($T
\sim 10^{11}$ K), and they may give off thermal radiation while
cooling. Pulsar activity may contribute to the warming, but after a
few tens of millions years the supply of energy, both rotational
and internal, will be essentially exhausted. From then on, NSs are
cold, dead objects, apart for some sudden release of internal
energy, like seismic activity, which has been often proposed but
never convincingly observed. Since the Galaxy is  $\sim 10^{10}$ yr
old, one infers that the active state of NSs is a tiny fraction of
their life, or, conversely, that the Galaxy is the graveyard of a
multitude of dead NSs.

The dowry of collapsed objects is their gravitational field, which
in the case of NSs is strong, since their radius is about 3
gravitational radii. NSs in the Galactic halo  may become
responsible for lensing episodes, and in fact they have been
invoked in that context
(see e.g. Wasserman \& Salpeter 1994\markcite{ws}). Also, 
many investigations suggest that the magnetosphere of old NSs
accreting from dense clouds may be bounded by a collisionless shock (see
e.g. Arons \& Lea 1976\markcite{al76}), although a
self--consistent analysis probing its existence has not been presented
yet. In this picture, the Alfv\`en radius could be a possible site
for acceleration of cosmic rays (Shemi 1995\markcite{sh95}).
Inelastic $p-p$ collisions
between the accelerated particles and the accreting flow may also produce
300~GeV--1~TeV $\gamma$--rays, that may be detected by Cerenkov
experiments
(Blasi 1996\markcite{b96}). However, despite these remain promising
possibilities, up to now no NS has been 
detected unambigously
through these techniques.

Dead neutron stars may, however, be brought to life again by
accretion from the interstellar medium (ISM). This idea, which is
the main guideline of this review, was proposed some 30 years ago
by Ostriker, Rees \& Silk (1970\markcite{ors70}, hereafter ORS),
but for two decades it remained latent in the
impetuous progress of X--ray astronomy. It was only in the
early '90s that Treves \& Colpi (1991\markcite{tc91}) and Blaes \&
Madau
(1993\markcite{bm93}), hereafter TC and BM,
realized that the ROSAT X--ray telescope (launched in May 1990)
could have the capability of detecting NSs accreting from the ISM.
These conclusions motivated a renewed interest in the field, and
several studies aiming to estimate the number of detectable
sources in the more favorable sites in the Galaxy were presented.

It has become
increasingly clear that a realistic estimate of the observability of old,
isolated, accreting neutron stars (ONSs) relies on a thorough
understanding of a number of issues, such us
\begin{mylist}{xx}
\item[$\bullet$] {the present density and velocity distribution of NSs,
a subject of interest in itself, strongly related to
pulsars birth rate and kick distribution. The accretion
rate is, in fact, inversely proportional to the star velocity cubed;}

\item[$\bullet$] {the magnetic field and spin evolution of isolated NSs,
an 
unsolved problem in compact object astrophysics. Their knowledge is
crucial for old neutron stars, since accretion onto a rotating
dipole depends critically on the field strength and the
rotation period;}

\item[$\bullet$] {the distribution of the interstellar medium (ISM) and
Giant
Molecular Clouds in the Galaxy. The ONS luminosity scales
linearly with the density of the ambient medium, so brighter sources are
found where the ISM is denser;}

\item[$\bullet$] {the spectral properties of radiation emitted by neutron stars
accreting at low rates, and their dependence on the magnetic
field. }
\end{mylist}
In all these fields there has been substantial progress over the last few
years.

Spectra from accreting NSs for very low
luminosities have been computed for both unmagnetized and magnetized
atmospheres by
Zampieri \etal (1995)\markcite{ztzt} and Zane, Turolla \& Treves
(1999)\markcite{mag}.
This regime was never explored
before, since attention was mainly focussed on X--ray
binaries, where the luminosity is not far from the Eddington
limit, millions of times that expected from ONSs accreting from the ISM
(see section~3).

The present distribution of NSs has been studied by several
authors (Pac\'zynski 1990\markcite{pac1990}; Blaes \& Rajagopal
1991\markcite{br91}; BM\markcite{bm93}; Zane \etal 1995\markcite{ons}) and
led to
the prediction that a large number of accreting ONSs should be
detectable with ROSAT in regions where the ISM is denser, such us the
closer Giant Molecular Clouds (BM; Colpi, Campana \& Treves
1993\markcite{cct}; Zane \etal 1995\markcite{ons}),  or in the
Solar vicinity (Zane \etal 1996b\markcite{solpr96}). It was also
suggested that the collective emission from the ONS population
could account for part of the galactic excess in the soft X--ray
background  (Zane \etal 1995\markcite{ons}) and for the soft component
of the diffuse X--ray emission from the Galactic Center (Zane,
Turolla \& Treves 1996\markcite{cgal}).

However, as soon as the first results of the searches for isolated
NSs were published (Motch \etal 1997\markcite{mo97}; Belloni,
Zampieri \& Campana 1997\markcite{bzc97}; Maoz, Ofek \& Shemi
1997\markcite{mos97}; Danner 1998a,
b\markcite{d98a}\markcite{d98b}), it became apparent that
theoretical estimates were in excess by a factor at least 10--100
over the upper limits implied by observations. Still, in the last
few years half a dozen objects found in ROSAT images appear as
good candidates for close--by ONSs accreting from the ISM 
(Stocke \etal 1995\markcite{sto95}; Walter, Wolk \& Neuh\"auser
1996\markcite{wwn96}; Haberl \etal 1996\markcite{ha96},
1997\markcite{ha97}; Haberl, Motch \& Pietsch 1998\markcite{mp98};
Schwope \etal 1999\markcite{sch99}; Motch \etal 1999\markcite{mo99}; 
Haberl, Pietsch \& Motch 1999\markcite{hpm99}).

The most recent line of research is therefore twofold. First it
aims to explain the paucity of observed ONSs. At present, the most
likely answers to this puzzling question are that NSs have a much
larger mean velocity than that assumed in the past (so the
luminosity is drastically reduced), or that the magnetic field
decays over a timescale $\approx 10^8-10^9$ yr (preventing the
star from entering the accretion stage). In both cases upper limits on
the number of observable ONSs can be used to constrain the
population properties (Colpi \etal 1998\markcite{elu98}; Livio, Xu
\& Frank 1998\markcite{lxf98}; Popov \etal 1999\markcite{pop99}).
The second goal is to provide a definite assessment of the nature
of the seven candidates proposed so far. Although their
identification with isolated NSs seems rather firm, it is still
unclear if they are rather young cooling objects, giving off
thermal radiation at the expense of their internal energy, or
accreting ONSs. Definite proof may come from observations at
optical wavelengths, where the two classes of sources should show
rather distinct emission properties (Zane, Turolla \& Treves
1999\markcite{mag}).

New generation X--ray satellites (CHANDRA, XMM, ASTRO E), may
either increase the number of ONS candidates, or lend further
support to their
paucity. Since these new missions are becoming operational, it seems
appropriate to review now our present knowledge of the subject.

The structure of the paper is the following. In section
\ref{general} we summarize the basic conditions under which accretion
is possible and section \ref{spectrum} contains a discussion about
the spectral properties. The conditons affecting the number 
of old, accreting NSs are
analyzed in section \ref{distribu}. Section \ref{theory} reviews
theoretical
investigations about the observability of single sources. 
A report of the present status
of observations is given in section \ref{observations} and the
observational appearance of accreting and cooling NSs is discussed
in section \ref{avsc}. In section \ref{paucity} possible
explanations for the paucity of accretors are considered. Our
conclusions follow in section \ref{conclu}.

\section{Interaction with the Interstellar Medium: Various Types of
NS}\label{general}

Let us consider a NS of mass $M$, radius $R$, spin period $P$,
magnetic field $B$, moving with velocity $V$ relative to an
ambient medium of number density $n$. There
are various possible modes of interaction between the NS and the
medium (see e.g. Davidson \& Ostriker 1973\markcite{do73}; Lipunov
1992\markcite{lipu}, and, specifically
for the case of isolated NSs, Treves, Colpi \& Lipunov
1993\markcite{tcl93}; Popov \etal 1999\markcite{pop99}) and these are well
illustrated by introducing
different characteristic lengths. The most relevant ones are the
accretion, the corotation and the Alfv\`en radius, all of them
familiar from the theory of binary X--ray sources.

The accretion radius, $r_{acc}$ defines the region where the
dynamics of the ISM is dominated by the gravitational field of
the NS and is given by \be\label{racc} r_{acc} = {2GM \over v^2} \sim
3\times 10^{14}\, m v_{10}^{-2} \ {\rm cm} \, , \ee where $v_{10}
= (V^2+C_s^2)^{1/2} /(10$ \kms), $m =M/M_\odot$, $M_\odot$ is the
solar mass and $C_s\sim 10$ \kmss is the ISM sound speed. Equation
(\ref{racc}) is derived in the framework of the Hoyle--Bondi
theory of accretion. The resulting value of $r_{acc}$ is only
approximate, but is sufficient for our purposes.

Similarly, the Alfv\`en radius, $r_A$, is the boundary inside
which the dynamics of the infalling matter is dominated by the NS
magnetic field. In spherical symmetry,
\begin{eqnarray}
r_A & =
& \left({{B^2R^6}\over{\sqrt{2GM}\mdot}}\right)^{2/7}\sim\nonumber\\
& \sim & 2\times 10^{10} B_{12}^{4/7}\mdot_{11}^{-2/7}R_{6}^{12/7}m^{-1/7}
\, {\rm cm} \end{eqnarray}
where $\mdot$ is the accretion rate (see eq. [\ref{mdot}]),
$B_{12}=B/(10^{12}$ G), $\mdot_{11} = \mdot/(10^{11}$ g s$^{-1}$) and
$R_{6} =R/(10^6$ cm).

Finally, the corotation radius is obtained by equating the angular
velocity of the NS with the Keplerian angular velocity,
\be
r_{co} = \left(\frac{GMP^2}{4\pi^2}\right)^{1/3}\sim 2\times 10^8
m^{1/3} P^{2/3} \, {\rm cm}\, . \ee

In the case of a NS, strong magnetic fields and fast rotation may
inhibit accretion because of the momentum outflow, produced by the
spinning dipole, and the corotating magnetosphere (ORS; Illarionov
\& Sunyaev 1975\markcite{is75}; Davies \& Pringle
1981\markcite{dp81}; Blaes \etal 1992\markcite{b92}). Three basic
conditions must be verified in order to make accretion possible.
First, accretion is inhibited if the Alfv\`en radius is
larger than the accretion radius in which case the system remains
in the so--called {\em georotator} stage.

Second, at the accretion radius the gravitational energy density of the
incoming material
\be
U_G = {{GMm_pn}\over r}\sim 6.5\times
10^{-13}\mdot_{11}r_{14}^{-5/2} \ {\rm erg\, cm}^{-3} \ee must be
greater than the energy density, $U_B$, of the relativistic
momentum outflow produced by the rotating $B$--field. If the field
is dipolar,
\begin{eqnarray}
U_B&=&\left({{B^2}\over{8\pi}}\right)\left({{R^6}\over{r_c^6}}\right)\left(
{{r_c^2}\over{r^2}}\right) \sim \nonumber\\ &\sim& 7.5\times
10^{-9}B_{12}^2P^{-4}R_{6}r_{14}^{-2} \ {\rm erg\, cm}^{-3}\, ,
\end{eqnarray} where $r_{14} = r /(10^{14}$ cm) and $r_c =
cP/2\pi$ is the light cylinder radius. This condition is met only when the
NS has spun down to a period \be P \gtrsim P_{crit}\sim 10\,
B_{12}^{1/2}\mdot_{11}^{-1/4}(r_A)_{14} R_{6}^{3/2} m^{-1/8} \
{\rm s}\, , \ee while during the time in which $ P < P_{crit}$ the
system is in the {\em ejector} phase, with no accretion occurring.
The duration of this phase may largely exceed the pulsar life time
($\sim 10^7$ yr).

Since the star is slowing down at the magnetic dipole rate, this
first barrier is overcome in a typical time scale $t_1 \sim 4
B_{12}^{-1} \dot M_{11}^{-1/2} \ {\rm Gyr}$. As noted by BM, this
value is uncomfortably close to the age of the Galaxy. However, if
taken literally, it implies that a large fraction of ONSs  can
have spun down sufficiently, at least if the majority of them are
born early in the Galactic history.

After $P$ has increased above $P_{crit}$, the infalling material proceeds
undisturbed until the Alfv\`en radius, where the NS magnetic energy
density balances the matter bulk kinetic energy density. The
corotating magnetosphere will then prevent the accreting material from
going
any further, unless the gravitational acceleration at the Alfv\`en radius
is larger than the centrifugal pull, i.e.
\be
{{GM}\over{r_A^2}}\gtrsim \left({{2\pi}\over P}\right)^2r_A\, .
\ee This translates into another stronger constraint on the value
of the period, since it must be \be\label{psec} P\gtrsim P_A\sim
10^3B_{12}^{6/7}\mdot_{11}^{-1/2}m^{-1/2} \ {\rm s}\, , \ee (our
third condition) otherwise matter will accumulate at the Alfv\`en
radius and the system remains in the so--called {\em propeller}
phase.

The centrifugal barrier at the Alfv\`en radius poses a severe
problem, since $P_A$ is so large that it cannot be reached only by
means of magnetic dipole radiation. However, there are at least two other
effects that may play an important role, making accretion
possible: the decay of the $B$--field and the torque exerted by
the accreting material on the NS itself.
The accreted matter will be spun up by the magnetosphere, and it
will exert a torque on the NS (see e.g. BM).
The propeller physics is very complicated, and a detailed review of this
issue is outside the scope of this paper. Its thorough understanding
requires a full 2--D or even 3--D MHD investigation of the interaction
between accretion flow and rotating magnetosphere (see e.g. 
Toropin \etal 1999\markcite{to99}). Nevertheless, approximated
expressions for the
torque were presented by BM (see also, for other spin--down formulae,
Lipunov \& Popov 1995\markcite{lp95}).
Based on that treatment, the
corresponding spin--down time to $P_A$ turns out to be $\sim 0.04
B_{12}^{-11/14} n^{-17/28} v_{10}^{29/14}$ Gyr, a value adequate
to allow interstellar accretion even for high values of the
magnetic field. However, for such high fields
accretion is probably unsteady and ONSs might
appear as transient X--ray sources (Treves, Colpi \& Lipunov
1993\markcite{tcl93}).
Numerical simulations indicate that more rapid spin down occurs either
because of the large mass expulsion 
rate (Toropin \etal 1999\markcite{to99}; 
Toropin, private communication)
or because the
material builds up, compressing the magnetosphere and becoming unstable to
large scale mixing with the $B$--field (Wang \& Robertson
1985\markcite{wr85}).

In any case, if the second barrier is also overcome, then matter
flows down onto the NS surface and the system becomes an {\em
accretor}. Eventually, at least in the simplest picture, if the
magnetic field is strong enough to channel the accretion flow, the
gas slides along the open field lines and the emitting region is
restricted to two polar caps of radius
\be
r_{cap}\sim {{R^{3/2}}\over{r_A^{1/2}}}\sim 7\times
10^3B_{12}^{-2/7} \dot M_{11}^{1/7}R_6^{9/14}m^{1/14} \ {\rm cm}\,
. \ee

The accretion rate is a sensitive function of the star velocity
and, again within the Bondi--Hoyle theory, is given by
\be
\label{mdot}
\dot M= \frac{2\pi (GM)^2 m_p n}{(V^2+C_s^2)^{3/2}}\simeq
10^{11}\, n\, v_{10}^{-3}\ {\rm g\, s}^{-1} \, . \ee
The total luminosity is then
\be
\label{lu} L\sim \frac{GM}{R}\dot M \sim 2 \times 10^{31}\dot
M_{11}
 \,  {\rm erg \, s^{-1}}
\ee which is orders of magnitude below the Eddington limit. This
implies that matter is very nearly in free--fall until it reaches
the NS surface. The dynamical time is then much shorter than the
radiative cooling times and virtually no energy is released before
the flow hits the outermost stellar layers. Here accreting protons
are decelerated by Coulomb collisions with atmospheric electrons
and/or by plasma interactions and the flow stops after penetrating
a few ($\lesssim 10$) Thomson depths in the NS atmosphere. The
bulk kinetic energy of the infalling protons is transformed into
thermal energy and finally converted into electromagnetic
radiation.

\begin{figure}
\epsfxsize=\hsize
\centerline{{\epsfbox{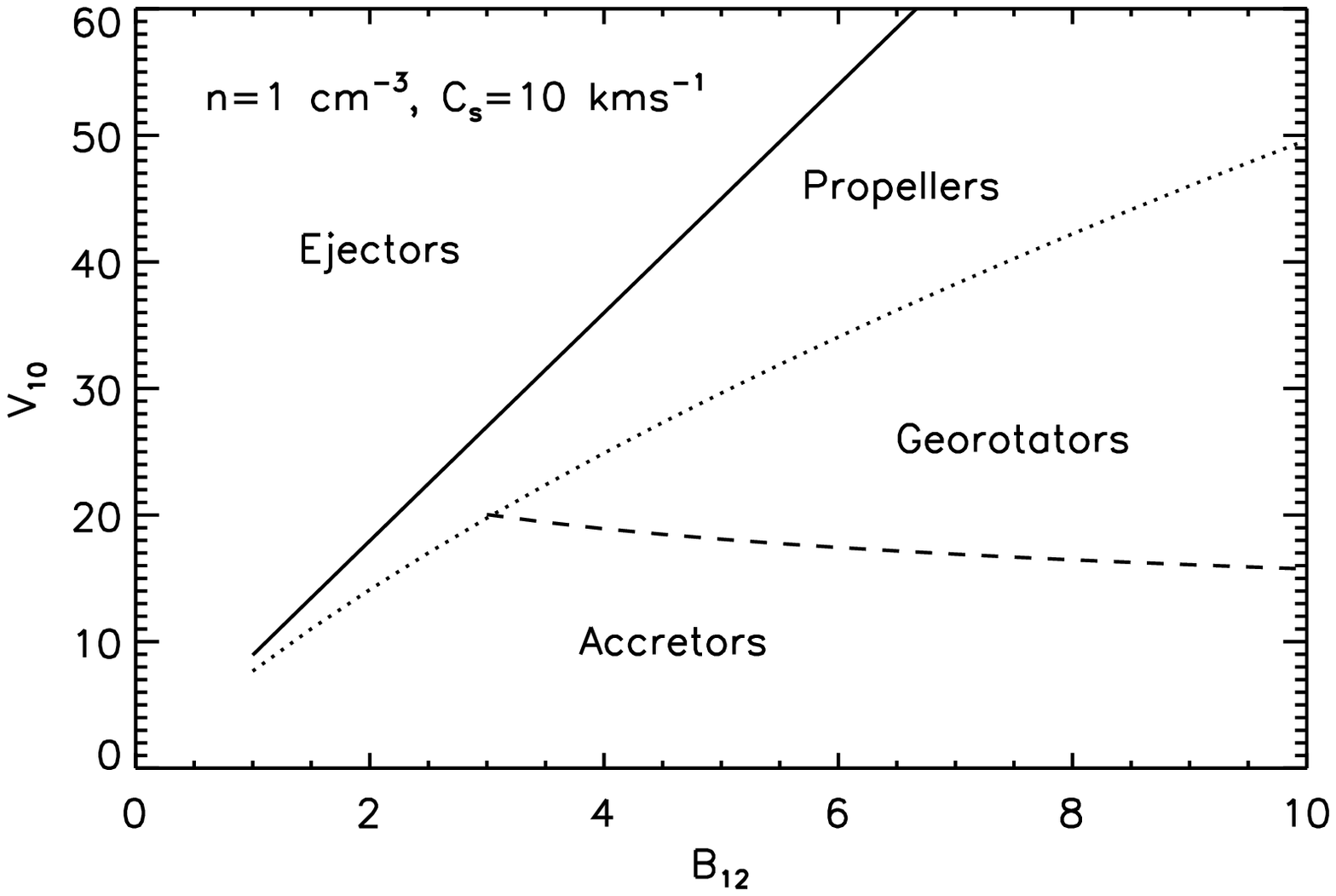}}}
\caption{The different stages of an old neutron star as a function of the star
velocity, in units of $10 \ {\rm km\, s}^{-1}$, and magnetic field,
in units of $10^{12}$ G.
\label{stages}}
\end{figure}

An example of the possible stages for a NS 10 Gyr old as a function of the 
star velocity and magnetic field is shown in figure \ref{stages}.

\section{The Spectrum of Accreting NS}\label{spectrum}

Due to the combined effects introduced by the response of the
detector and absorption by the ISM, theoretical estimates on
ONS observability and the choice of the most favorable energy
bands for their detection are crucially related to the
determination of the spectral properties of the emitted radiation
and to the evaluation of the mean photon energy in particular.

\subsection{Black Body Spectrum }\label{bbspec}

Many investigators assumed (and still do!) that the spectrum
emitted by NSs accreting from the ISM is a blackbody at the star
effective temperature,
\begin{eqnarray}
T_{eff}&=&\left({L\over{4\pi
fR^2\sigma}}\right)^{1/4}\sim\nonumber\\ & \sim & 3.4\times 10^5
L_{31}^{1/4}R_6^{-1/2}f^{-1/4} \ {\rm K}\, , \end{eqnarray} where
$f$ is the fraction of the star surface covered by accretion and
$L_{31} = L / (10^{31} \, {\rm erg \, s^{-1}})$ (see eq.
[\ref{lu}]). Although crude, such approximation is basically
correct, since the density in the deep atmospheric layers for
$T\sim T_{eff}$ turns out to be

\be \rho\approx{{GM m_p \tes}\over{R^2kT_{eff}}}\sim 16 \tes m
R_6^{-2} (T_{eff})_5^{-1} \ {\rm g \, cm^{-3}} \ee where $
(T_{eff})_5 = T_{eff}/(10^5 \, {\rm K})$. For these values of
density and temperature, the free--free optical depth
$\tau_{ff}\approx (\kff/ \kes )\tes $ is much larger than unity up
to $h\nu\gg kT_{eff}$, so thermal equilibrium is established.
Compton scattering is not expected to modify the spectrum because
of the relatively low Thomson depth and electron temperature. Cold
atmospheric electrons emit cyclotron radiation, but for $B\sim
10^9$~G the cyclotron line contribution to the total luminosity is
vanishingly small and it never exceeds a few percent even for
$B\sim 10^{12}$ G (Nelson \etal 1995\markcite{nel95}). These
considerations show that, at least to first order approximation, the
emission is peaked at an energy
\be
E\sim 3kT_{eff}\sim 100\, L_{31}^{1/4}R_6^{-1/2}f^{-1/4} \ {\rm
eV} \ee which falls in the soft X--ray range (see ORS).

For a low  magnetic field ($B\lesssim 10^{10}$ G)
emission/absorption processes (and radiative transfer) are much
the same as in the non--magnetic case, but the emitting area is
now substantially reduced because $r_{cap}\ll R$. This has an
important effect on the emitted spectrum, since, for a given
luminosity, the effective temperature increases with decreasing
$f$
\begin{eqnarray}
\label{teffmag} {{T_{eff}}\over{T_{eff,B=0}}}&=&
f^{-1/4}=\left({{4\pi R^2}\over{2\pi r_{cap}^2}}\right)^{1/4}\sim
\nonumber \\ &\sim &13\, B_{12}^{1/7}\dot
M_{11}^{-1/14}R_6^{5/28}m^{-1/28}\, .
\end{eqnarray}
Of course the same
argument about the reduced size of the radiating region holds also
for strongly magnetized NSs ($B>10^{10}$~G), although in the
latter case the different response properties of the magnetoactive
plasma must be accounted for.

The typical values of $T_{eff}$ and $L$ indicate the two basic
reasons that make the detection of these sources extremely
difficult: their intrinsic weakness (the luminosity is orders of
magnitude below Solar), and the fact that the their spectrum peaks
in the EUV--soft X--rays, an energy band which is strongly
absorbed and not easily accessible even to spaceborne
instrumentation. A definite prediction, which is largely
independent of the details of the spectral distribution, is that
the X--ray to optical flux ratio 
must be exceedingly
large
for
these sources,
\be
\log \left(\frac{f_X}{f_V}\right)\sim 5.5
+3\log\left(\frac{kT_{eff}} {100\, {\rm eV}}\right) \ee for a
blackbody spectrum, where $f_X$ and $f_V$ are the fluxes integrated
in the intervals 
0.5--2.5 keV, and 3,000--6,000 A, respectively.

\subsection{Beyond the Black Body Spectrum}\label{trspec}

The problem of determining the detailed radiation spectrum from
accreting NSs was first addressed by Zel'dovich \& Shakura
(1969\markcite{zs69}) and further investigated by Alme \& Wilson
(1973\markcite{aw73}). Both authors considered a static,
plane--parallel, pure H atmosphere with negligible magnetic field,
kept hot by the energy released by the infalling protons, and
focussed on the high accretion rates, typical of X--ray binaries.
Models are characterized by two basic parameters, the accretion
rate and the penetration depth of protons falling into the NS
outer layers. Using the same input physics, 
Zampieri \etal (1995)\markcite{ztzt}
extended the calculations to the low luminosity range ($L\gtrsim
10^{30}$ erg/s) relevant to accretion from the ISM. They  found
that the emerging spectrum is harder with respect to a blackbody
at $T_{eff}$. The hardening factor $T_\gamma/T_{eff}$, where
$T_\gamma$ is the radiation temperature, is 2--3 and increases for
decreasing luminosity. Not surprisingly, the same behaviour is
shared by (unmagnetized, H) atmospheres around cooling NSs (Romani
1987\markcite{ro87}). In fact, the hardening is not related to the
heat source but to the frequency dependence of the free--free
opacity (the most important radiative process). Higher energies
decouple in deeper layers, which are hotter, and the final
spectrum is, roughly, the superposition of blackbody spectra at
different temperatures.

\begin{figure}
\epsfxsize=\hsize
\centerline{{\epsfbox{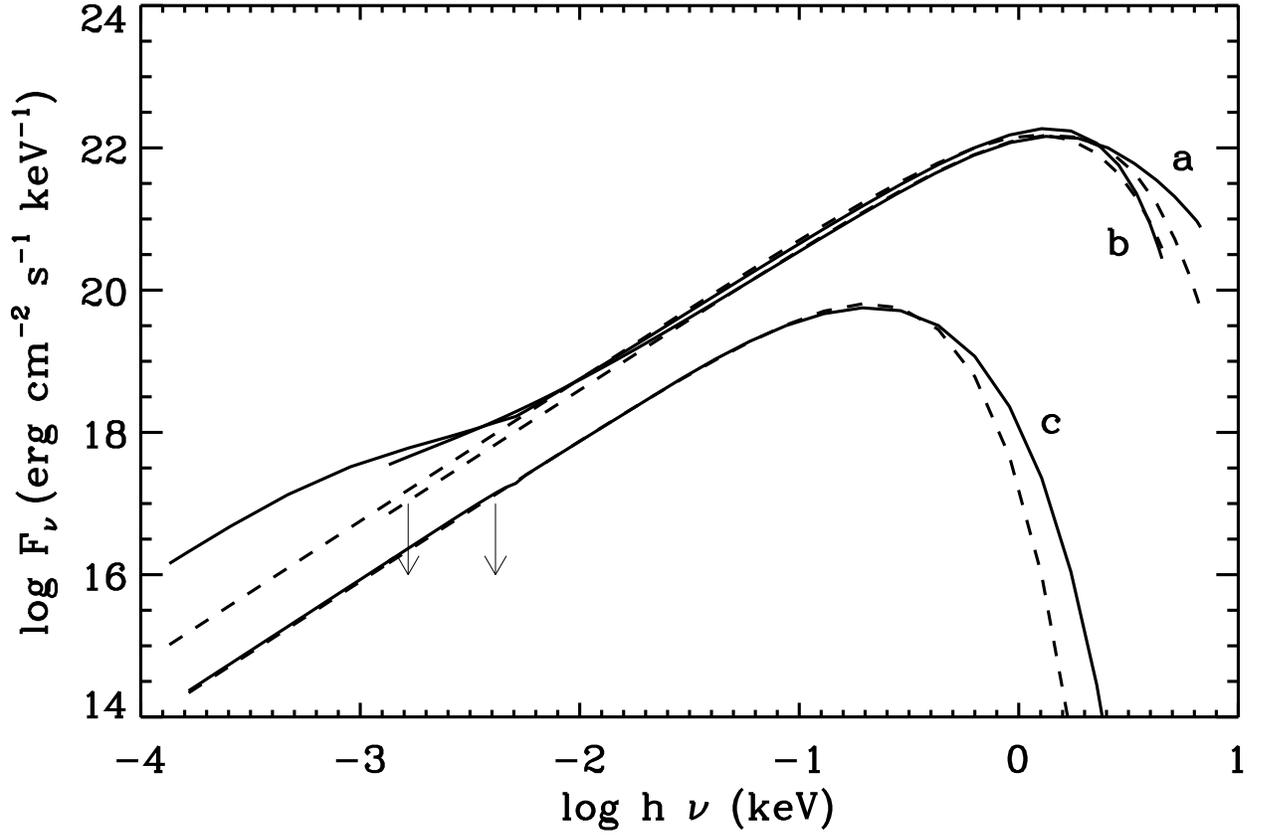}}}
\caption{Synthetic spectra of pure H accreting (upper curves) and cooling
(lower curve) NSs. a): $B = 0$~G, $L = 4.3 \times 10^{33}$ erg/s; b): $B =
10^{12}$~G, $L = 3.7 \times 10^{33}$ erg/s; c): $B = 10^{12}$~G, $L = 0.2
\times 10^{33}$ erg/s. The blackbody
spectra which best--fit the X--ray range close to the peack of each
spectrum are also shown (dashed curves).
The two arrows mark the optical band (3000--7500 A).
\label{spectra}}
\end{figure}

A detailed computation of spectra emerging from strongly
magnetized, accretion atmospheres has been recently presented by
Zane, Turolla \& Treves (1999)\markcite{mag}. Extending previous
work, (see e.g. M\'esz\'aros 1992\markcite{mes92} and references
therein) they solved the transfer equations in a slab geometry for
the two normal modes in the (anisotropic) magnetoactive medium
coupled to the hydrostatic and energy balance equations. They have
shown that, for $L\sim 10^{30}- 10^{33}$ erg$\, \rm s^{-1}$, the
hard tail present in non--magnetic models with comparable
luminosity is suppressed and the X--ray spectrum, although still
harder than a blackbody at $T_{eff}$, is nearly planckian in
shape. At X--ray energies the spectral distribution closely
resembles that of a cooling, hydrogen atmosphere with similar
properties (see e.g. Shibanov \etal 1992\markcite{sh92} and
section \ref{avsc}). However, the two models (accreting/cooling)
differ rather substantially at optical wavelengths, where the
former exhibits a definite excess over the Rayleigh--Jeans tail of
the X--ray best--fitting blackbody (see figure \ref{spectra}). A
more thorough comparison of the properties of accreting and
cooling models, and of the observational relevance of the optical
excess as a signature of accretion, is postponed to section
\ref{avsc}.

In concluding this section, we remark that, although this has
little effect on the bulk of the emission, for $B\sim 10^{12}$ G
the (broad) electron cyclotron line, centered at $\sim 11.6\,
B_{12}$ keV, becomes a prominent spectral feature over the
falling--off Wien continuum, as pointed out by 
Nelson \etal (1995)\markcite{nel95}. The detection of the cyclotron line
in
conjunction with  polarization measures near the proton cyclotron
energy (where the degree of polarization crosses zero) may prove a
powerful tool in determining the field strength, even in the
absence of pulsations (Zane, Turolla \& Treves 1999\markcite{mag};
Pavlov \& Zavlin 1999\markcite{pz99}).

\section{Conditions Affecting the Number of Accreting ONS in the 
Galaxy}\label{distribu}

\subsection{The Number of NSs in the Galaxy}

The total number of neutron stars in the Galaxy is still
uncertain, varying between $N_{tot}\sim 10^8$ and $\sim 10^9$.
These figures are inferred from constraints on the nucleosynthetic
yields produced in core collapse, and from estimates on the
present rate of Type II supernova events, as observed in nearby
spiral galaxies.

According to Arnett, Schramm \& Truran (1989\markcite{ar89}), core
collapse events occur at a rate of one every 10 years and, when averaged
over the entire Galaxy lifetime, can account for nearly half of the
Galactic Fe abundance. This would imply $\sim 10^9$ NSs if the
disk of the Milky Way has a mass of $\sim 10^{11}M_{\odot}$ and an
age of $\sim 10^{10}$~yr. Since  thermonuclear events involving
the explosion of C/O white dwarfs may contribute to the Fe yield,
$N_{tot}$ cannot be determined with higher accuracy.

The present rate of Type II supernova explosions, derived from
Capellaro \etal (1997\markcite{cap97};  see also Madau, Della
Valle \& Panagia 1999\markcite{mad98}) is $\sim 0.5\, h_{50}^2$
SNu, where 1 SNu is 1 supernova per 100 yr per $10^{10}L_{B\odot}$ and
$h_{50} $ the Hubble constant in units of 50 \kms Mpc$^{-1}$. 
For a Galactic blue luminosity of $5\times
10^{10}L_{B\odot}$, this estimate implies a supernova event every
30 years, approximately. Thus, $N_{tot}\sim 3 \times 10^8$ NSs
should be present in the Milky Way. However, this estimate may
just give a lower limit for $N_{tot}$ since the Galaxy may have
experienced phases of enhanced star formation at early epochs. The
rate in the past could have been higher by a factor of 10 (Madau
\etal 1996\markcite{mad96}) yielding again $N_{tot}$ close to
$10^9$.

\subsection{Space and Velocity Distribution of
NSs}\label{distribuv}

The visibility of accreting ONSs is severely constrained by the
present velocity distribution, in particular by the extent of a
low velocity tail. As old NSs outnumber young objects, in all
likelihood active as pulsars, potential accretors belong
predominantly to the dynamically old population which had the time
to evolve in the Galactic potential. The velocity distribution of
ONSs and their spatial density in the volume accessible to
observations can be inferred simulating a large, statistically
significant number of orbits in the Galaxy. The strategy followed
by a number of authors (BM; Blaes \& Rajagopal 1991\markcite{br91};
Frei, Huang \& Paczy\'nski 1992\markcite{fhp92}; Zane
\etal 1995\markcite{ons}) is to start with  a seed population,
born in the disk with natal kicks determined according to a
specific velocity distribution, that evolves dynamically in the
Galactic potential. Customarily the distribution of the kick
velocity is approximated as isotropic and Gaussian. The neutron
star birth rate  and spatial distribution mirror those of massive
stars in the Galaxy,  while pulsars, representative of the parent
population, are used to constrain the natal velocity dispersion.

The first to use available velocity information in the statistical
analysis of young NSs were Narayan \& Ostriker
(1990\markcite{no90}). They found that observations of  periods
and magnetic fields from  a sample of about 300 pulsars are well
fitted by invoking the presence of two different populations of
NSs at birth, slow (S) and fast (F) rotators. Both  have an
isotropic Gaussian distribution (relative to the circular speed),
the F rotators being  characterized by a mean velocity  $V_F \sim
60 $ \kms, and by a scale height $z_F\sim 150$ pc, much lower than
the S ones for which $V_S \sim 240 $ \kmss and  $z_S = 450$ pc. From this
fit, the evolved distribution function of ONSs has been
obtained by Blaes \& Rajagopal (1991\markcite{br91}), BM, and Zane
\etal (1995\markcite{ons}). As far as accretion onto ONSs is
concerned, only the F population, which represents $\sim 55$ \% of
the total number, is relevant. Results from Zane \etal 
(1995\markcite{ons}) show that, after secular evolution, in the
local region $7.5 \ {\rm kpc} \leq R \leq 9.5 \ {\rm kpc}$, ONSs
are characterized by a mean scaleheight  $\langle z \rangle \sim
250$ pc and by a mean velocity, averaged over $|z| \leq 200$ pc,
$\langle V \rangle \sim 78 \ {\rm km \, s^{-1}}$. The number
density of NSs, within 2 kpc from the Sun, turns out to be $ n_0
\sim 3\times 10^{-4}(N_{tot}/10^9) \, {\rm pc^{-3}}$.

The previous scenario was challenged by Lyne \& Lorimer
(1994\markcite{ll94}), who suggested that NSs acquire at birth
velocities significantly higher than those of both the F and S
sub--populations of Narayan \& Ostriker\markcite{no90}, the
imprint of an anisotropic core collapse during a type II
supernova. These authors derived from the observations a typical
mean  velocity  of about 450 \kms. Recently, on completion of a
quantitative analysis, the kinematics of radio pulsars was
assessed using population synthesis models and a sample of pulsars
with improved statistics and revised distance scale. Yet, the
results depicting the true pulsar population are still not
univocal, due to different criteria used in selecting the
sample and in treating observational errors. While Lorimer, Bailes
\& Harrison (1997\markcite{lbh96}) find  a mean natal velocity of
500 \kms, Hansen \& Phinney (1997\markcite{hp97}) claim a lower
value, $\sim 250-300$ \kms. Cordes \& Chernoff
(1998\markcite{cc98}) suggest a possible decomposition of the
velocity distribution into two components with mean velocities of
197, 700 \kmss and fractions 0.84, 0.16, respectively. Despite
these controversies, recent observations seem  to point
towards an increase in the mean velocity of young NSs relative to
Narayan \& Ostriker.

Accreting NSs belong  necessarily to the low velocity tail of the
population, and at present the low statistics prevents a
determination of the extension of such a tail with the required accuracy.
Hartmann (1997\markcite{h97}), Hartmann \etal (1997\markcite{hal97}) and
Iben \& Tutukov (1998\markcite{it98}) do
not exclude the presence of a prominent low velocity component.
Dynamical heating (Madau \& Blaes 1994\markcite{mb94}) may influence the
evolution of slow NSs. This process, observed in the local disk
stellar population, causes the velocity dispersion to increase
with time as a consequence of scattering by molecular clouds and
spiral arms (Wielen 1977\markcite{w77}). If ONSs participate in the same
process,
dynamical heating over the lifetime of the Galaxy may scatter a
fraction of low velocity stars to higher speeds and this, in turn,
could devoid the distribution of  slow stars.

\subsection{The Distribution of the Interstellar Medium}\label{ism}

The discussion of the properties of accreting sources, and the
study of the detectability of ONSs, require a detailed mapping of
the interstellar medium. For a given source, the interstellar
density determines both the intrinsic luminosity and the amount of
absorption along the line of sight. The situation is more
uncertain and complex when dealing with particular regions such as the
center of the Galaxy or the Solar proximity, while the gas
structure is smoother on larger scales and numerical fits to the
gas distribution have been published in the literature (see e.g.
Dickey \& Lockman 1990\markcite{dl90} for a detailed review).

In particular, observational data from Ly$\alpha$ and 21 cm absorption
measures show that the ISM distribution of both cold and warm HI is
nearly constant in radius while its $z$--dependence can be fitted by
\be
\label{nh} n_{HI} = \sum_{i=1}^2 n_i \exp {\left( -{ {z^2} \over { 2
\sigma_i^2
} } \right)} + n_3 \exp{\left( -{z \over h }\right)} \ee with $n_1 =
0.395$, $n_2 = 0.107$, $n_3 = 0.064 $; $\sigma_1 = 212$, $\sigma_2
= 530$ and $h = 403$ pc; $n_i$ and $\sigma_i$ are in cm$^{-3}$ and
pc, respectively. The applicability of the previous expression is
restricted to the range $0.4 \leq R/R_0 \leq 1$, where $R$ is the
galactocentric radius and $R_0 = 8.5 $ kpc is the distance of the
Sun from the Galactic center. The gas layer has a scale height of
about 230 pc in the vicinity of the Sun, while for $R \leq 0.4R_0$
it shrinks to $\approx 100$ pc and in the outer Galaxy it expands
linearly up to $\approx 2$ kpc. The other important contribution
to the total ISM density comes from molecular hydrogen. The best
tracer of H$_2$ is the CO molecule, and observational data suggest
a local gaussian distribution with a scale height of $\sim 60-75$
pc. Observations, however, are much less conclusive as far as the
midplane density is concerned (Bloemen 1987\markcite{b87}), also
because it may significantly depend on $R$ (De Boer
1991\markcite{db91}). As a first approximation, the H$_2$
distribution can be modeled as a gaussian with central density 0.6
${\rm cm ^{-3}}$ and FWHM 70 pc (De Boer 1991\markcite{db91}). The
ionized component gives only a very small contribution to the
total density, and can be neglected.

The evidence for a complex sub--parsec structure in the ISM has been
recently found by a number of authors on different scales, from 
$10^4-10^6$ AU down to $10-100$ AU (see e.g. Meyer \& Lauroesh
1999\markcite{ml99} and references therein). On all the sampled scales
observations imply dense concentrations of atomic gas ($n_H  \gtrsim  10^3
\, 
{\rm cm ^{-3}}$) in otherwise diffuse sight lines. 
These local condensations can not be accounted for in the standard McKee
\& Ostriker (1977\markcite{mo}) equilibrium pressure model for the ISM,
thus further 
interstellar line mapping is required to increase the small scale sky
coverage and to probe the sophisticated spatial structure. On the other
hand, their existence could explain why some ONS candidates are found
in very low average density regions.

Even leaving aside the issue of the ISM clumpiness on small scales, the
structure of the ISM at distances $\lesssim $
1 kpc the ISM is highly anisotropic and its complex morphology and
physical state are currently a very active field of research. At
the same time, this represents a particularly favorable site,
containing a high number of ONSs that, due to their vicinity,
may be relatively easy to detect. Assuming a spatial
density of $\sim 3\times 10^{-4} \ {\rm pc}^{-3}$ (section 4.2,
BM, Zane \etal 1995\markcite{ons}),
 about 150 ONSs are in fact expected in a sphere of radius 50 pc centered
on the Sun. Quite disappointingly, the observability of these
close--by NSs as accretion powered sources is severely hindered by
the shortage of fuel. In fact, the Sun is surrounded by a region,
the Local Bubble, where the plasma has both very low density ($n
\sim 0.05-0.07 \, {\rm cm }^{-3}$) and high temperature ($T \ga
10^5 $ K). In the scenario proposed by McKee \& Ostriker
(1977\markcite{mo}), see also Cox \& Anderson (1982\markcite{ca}),
Cox (1983\markcite{co83}), the hot gas would fill $\sim 70-80 \%$
of the interstellar space and a large number ($\sim 2 \times
10^4$) of cool ($T \sim 80$ K), roughly spherical clouds are
expected to be present. Observational data support this model for
the region beyond $\sim 50-100$ pc from the Sun (Knude
1979\markcite{knu79}), but, as discussed by Paresce
(1984\markcite{p84}), soft X--ray, radio and color excess surveys
seem to indicate that no clouds are present at smaller distances
and that the denser material is more probably organized into
large, elongated, moving fronts within $\sim 50$ pc. The Sun
itself is embedded in a medium, the Local Fluff, which is warm ($T
\approx 10^3-10^4$ K) and slightly overdense ($n \sim 0.1$
cm$^{-3}$) with respect to the Local Bubble on scales $\la 20$ pc
(Diamond, Jewell \& Ponman 1995\markcite{djp95}).

The present picture indicates that the contour of
the Local Bubble
in the Galactic plane is highly asymmetric, with four major
discontinuities in four different Galactic sectors (Paresce
1984\markcite{p84}). In particular, a wall of neutral hydrogen is
located very close to the Sun in the second quadrant, $ 15^\circ <
l < 120^\circ$. According to Paresce (1984\markcite{p84}), the
wall is roughly parallel to the $l = 330^\circ-150^\circ$ axis and
is located at $d \leq 16$ pc, with an estimated depth of about 35
pc. The $N_{HI}$ contours presented by Frisch \& York
(1983\markcite{fy83}) are generally farther away, with the denser
material ($n \sim 1$ cm$^{-3}$) at $\sim 90$ pc from the Sun.
Welsh \etal (1994\markcite{w94}) have derived a highly asymmetric
contour of the Local Bubble that in the second quadrant is roughly
intermediate between those presented by Paresce and Frisch \&
York. The minimum radius of the local cavity has been estimated to
be $\sim 25-30$ pc, but, as stressed by the same authors, their
indirect method could produce an underestimate of $N_{HI}$ at
distances smaller than 50 pc. An  analysis of ROSAT EUV
data (Diamond, Jewell \& Ponman 1995\markcite{djp95}) has shown
that $n$ reaches $\sim$ 1 cm$^{-3}$ at $\sim 25-30$ pc and this
result seems to be in agreement with the asymmetric contour found
by Welsh et al. more than with those of Paresce, Frisch \& York
(see also Pounds et al. 1993\markcite{pou93}).

\section{Are ONS Detectable ?}\label{theory}

As we have seen in the previous sections, the expected luminosity
from an ONS accreting from the ISM is $\lesssim 10^{31}$ erg$\,
\rm s^{-1}$ with a mean photon energy in the interval $\approx
0.1-1$ keV. The expected flux at Earth is $\lesssim 10^{-11}
(d/100\, {\rm pc})^{-2}\ {\rm erg\, cm^{-2}\, s^{-1}}$ (where $d$
is the distance) which translates into $\lesssim 1$ ROSAT PSPC
count$\, \rm{s}^{-1}$, well above the typical threshold of the
ROSAT All Sky Survey (RASS, $\sim 0.02$ count$\, \rm{s}^{-1}$).
The sensitivity of the ROSAT PSPC, together with its unprecedented
capability in the soft X--ray band ($\sim 0.1-2$ keV) put
accreting ONSs within observational reach. At the same time the
ROSAT HRI had a very high angular resolution, which is decisive 
for follow--up pointed multiwavelength observations of detected
sources.

\subsection{Selection Criteria for isolated NSs}

Since accreting ONSs should be detectable with ROSAT,
it is important to establish the most relevant selection criteria.
The basic ones, mostly independent of
details about the spectral shape, were already proposed by TC\markcite{tc91}:

\begin{mylist}{xx}
\item[$\bullet$]{the X--ray to optical flux ratio should be extreme 
(see section \ref{bbspec}), and the source should show no emission in
other
energy bands (radio, IR, $\gamma$--rays);}
\item[$\bullet$]{the X--ray spectrum should be thermal and soft,
with typical temperatures $T\approx 100$ eV (see section
\ref{bbspec});}
\item[$\bullet$]{the source must be weak, $L \lesssim 10^{31}$ erg/s;}
\item[$\bullet$]{the low luminosity makes only close--by sources
($d \approx 100$ pc) potentially detectable. This, in turn,
implies that the column density $N_H$ must be $ \lesssim
10^{21}\, {\rm cm^{-2}}$;}
\item[$\bullet$]{since $\dot M \propto n$, the spatial distribution of 
sources should correlate with the denser regions of the ISM;}
\end{mylist}

For close--by sources the visual magnitude should be $\sim$ 25--26 and
this might make both
the proper motion and the parallax measurable, especially if isolated NSs are
fast objects. Would the
velocity distribution be rich in slow stars, more distant sources,
that may appear in Deep Exposures,
must concentrate along the Galactic plane because the scale height of both
gas and ONSs is $\sim$ 200--300 pc.

\subsection{Number of Detectable Sources with ROSAT}\label{oldest}

The observed count rate for a
source of monochromatic luminosity $L_\nu$ is given by
\be
CR = \int_0^\infty{{L_\nu}\over{4\pi d^2}}\exp{-(\sigma_\nu
N_H)}A_\nu {{d\, \nu}\over{h\nu}} \ {\rm count\, s^{-1}} \ee where
$\sigma_\nu$ is the interstellar absorption cross section and
$A_\nu$ is the detector effective area. If $CR$ represents the
detector sensitivity limit, the previous expression can be used to
find the maximum distance, $d_{max}$, at which a star of given
luminosity $L$ produces a count rate above the threshold. The
total number of detectable ONSs in a given field is obtained by
integrating the NS distribution function, $f(r,l,b,v)$, over the
solid angle subtanding the field and out to $d_{max}$
\be
N = \int \, dv\int_\Omega\, d\Omega\int_0^{d_{max}}f(r,l,b,v)\,
dr\, ,
\ee
where $l,\, b$ are the usual Galactic coordinates.

Using the distribution function of Paczy\'nski (1990\markcite{pac1990}),
TC\markcite{tc91} concluded that $\sim~6000N_9$
sources, located within
500 pc, should appear in the RASS. They assumed polar cap
accretion ($B=10^9$ G), a blackbody spectrum and
considered a low value for the uniform ISM density ($n=0.07 \ {\rm
cm}^{-3}$), appropriate to the local bubble. A more thorough
investigation of ONS detectability has been presented by BM\markcite{bm93}. 
They followed the evolution of the population by
integrating the orbits of $10^5$ stars moving in the Galactic
potential over the Galaxy lifetime. Newly born NSs were assumed to
belong to the F sub--population of Narayan \& Ostriker
(1990\markcite{no90}). In
the solar proximity the present ONS mean velocity was found to be
$\sim 80$ \kmss and the fraction of stars with $V\lesssim 40$ \kms
$\sim 20\%$, two times lower than what is assumed by TC. Their
results were, nevertheless in substantial agreement with those of
TC and predicted that several thousands sources should be present
in PSPC exposures. 
Results by BM were improved by Manning, Jeffries \&
Willmore (1996\markcite{mjw96}), who computed the number of possible
detections in the Rosat Wide Field Camera survey with a Monte Carlo
simulation. By taking 
into account effects of a varying magnetic field, NS mass, and exposure
times during the survey, they were able to predict a significant depletion 
in the predicted numbers of detections if the magnetic field has not
decayed to values $\lesssim 10^{10}$~G. 

Relatively high count rates are expected from ONSs accreting in
the denser regions closer to the Sun. Of course, these regions
must be large enough to contain a statistically significant number
of slow NSs which are the most luminous. In this respect, dense
molecular clouds, where the ambient density is $\sim 100$ times
higher than the average, provide  a very favorable environment for
observing ONSs. Estimates by BM, Colpi, Campana \& Treves
(1993\markcite{cct}) and Zane \etal (1995\markcite{ons}), using different
assumptions
about the
emitted spectrum, indicated that the closer clouds should harbor
some relatively luminous sources. A detailed analysis of the
observability of ONSs accreting in the closest overdense regions
of the solar neighbourhood has been presented by Zane \etal
(1996a\markcite{solpr96}). As they pointed out, although the local
interstellar
medium is underdense and relatively hot, it contains at least one
region, the Wall, in which the gas density is $\sim 1 \ {\rm
cm}^{-3}$. The Wall extends between $\sim 20$ and $\sim 50$ pc and
its angular size is $\sim 6000 \ {\rm deg}^2$. According to these
authors, about 10 ONSs are expected to be detectable in the Wall
directions and, due to their vicinity, these sources should appear at the 
relatively high flux limit of 0.1 counts/s in the PSPC survey.

Besides the detectability of individual sources, ONSs could also
reveal themselves through their contribution to the diffuse X--ray
background. This point, originally suggested by ORS, was addressed
by BM, who found that the integrated flux above 0.5 keV is
negligible under the hypothesis of blackbody emission. Zane \etal
(1995\markcite{ons}) have readdressed this issue, considering polar cap
emission and
using synthetic spectra, and concluded that magnetized,
accreting ONSs can contribute up to $\sim 20 \%$ of the unresolved
soft X--ray excess observed at high latitudes in the X--ray
background (see also De Paolis \& Ingrosso 1997\markcite{dpi97} for a
discussion of the possible contribution from NSs in the Galactic halo). 

Since the diffuse emissivity of accreting ONSs depends on their
spatial concentration and on the density of the ISM, the Galactic
Center, where both the star and gas density are very high, could
provide an excellent site for revealing ONS emission. The GC is a
well known source of diffuse emission, investigated by many
missions and in particular by GRANAT. ART--P data in the 2.5--30
keV band (Sunyaev, Markevitch \& Pavlinsky 1993\markcite{smp93};
Markevitch, Sunyaev \& Pavlinsky 1993\markcite{msp93}) show the presence
of an elliptical
diffuse source with different properties below and above $\sim 8$
keV. Assuming that ONSs represent $\sim 1\%$ of stars in the GC
and that their distribution in both physical and velocity space
follows that of low--mass stars, Zane, Turolla \&
Treves (1996\markcite{cgal}) were able to reproduce satisfactorily the
GRANAT data at the lower energies.

\section{ROSAT results: The Crude Reality of
Observations}\label{observations}

The search for old, isolated NSs in ROSAT images is a by--product
of ``complete'' surveys. Typically one defines a flux threshold and
lists all sources which are brighter than that. The threshold
should be such that the probability of confusion with the
background noise is small. The second step is the search for an
optical counterpart on the basis of positional coincidence and
peculiar colours. Possible optical candidates are then studied
spectroscopically to see if the source belongs to one of the known
classes of X--ray emitters. All sources for which no viable
optical counterpart was found are classified as NOIDs
(non--optically identified objects). Isolated NSs candidates, for
which no counterpart is expected down to a visual magnitude $\sim
26$, must be looked for among NOIDs.

Generally NOIDs are further studied by reducing their X--ray error
box. The lack of obvious counterparts in archival
optical images, like the POSS, is used to set a lower limit on the
X--ray to optical flux ratio. Sources with  $f_X/f_V\gtrsim 1000$
are classified as potential isolated NS candidates. Obviously, the
number of NOIDs in a given region is an absolute upper limit on
the number of ONSs.

RASS and deep field exposures of the most favorable sites (such as GMCs)
have been searched for accreting ONSs and, although the
various surveys were conducted with rather inhomogeneous criteria,
an important result soon became apparent. The upper limits on ONS
candidates were substantially below expectations estimated by, at least, 
one order of magnitude.
Results from the surveys are reviewed in section \ref{surveys} and
the implications of this (negative) conclusion are examined in
section~\ref{paucity}.

Despite ONSs being much more elusive than originally expected, the
observational efforts of the last few years have produced some
promising candidates and their number keeps increasing: one in
1995, two in 1996, one in 1998 and three in 1999. All of them were
discovered, or re--discovered, as in the case of MS 0317.7-6647
and RXJ 185635-3754, with ROSAT, both serendipitously and in
survey data (mainly the ROSAT Bright Survey, RBS). The main
properties of the seven candidates discovered so far are reviewed in
the following subsections and summarized in
table~\ref{candidates}.
\begin{deluxetable}{lcccccl}
\tablecolumns{7} \tablewidth{0pc} \tablecaption{Isolated NS
candidates\label{candidates}} \tablenum{1} \tablehead{
\colhead{Source}& \colhead{PSPC} & \colhead{$T_{bb}$} &
\colhead{$N_H$} & \colhead{$\log f_X/f_V$} &\colhead{Period}
&\colhead{Refs.}\\
\colhead{} & \colhead{count$\, \rm{s}^{-1}$} & \colhead{eV} &
\colhead{$10^{20}\, {\rm cm}^{-2}$} & \colhead{} &\colhead{s} & \colhead{} }
\startdata
 MS 0317 & 0.03 & 200 & 40 & $> 1.8$ & -- & a\\
 RX J1856 & 3.64 & 57 & 2 & 4.9 & -- & b, c, d, e \\
 RX J0720 & 1.69 & 79 & 1.3 & 5.3 & 8.37 & f, g, h\\
 RBS1223 & 0.29 & 118 & $\sim 1$ & $> 4.1$ & -- & i\\
 RBS1556 & 0.88 & 100 & $< 1$ &$ > 3.5$ & -- & i, l\\
 RX J0806 & 0.38 & 78 & 2.5 & $> 3.4$ & -- & m\\
 RX J0420 & 0.11 & 57 & 1.7 & $>3.3$ & 22.7 & n\\
\tablecomments{a=Stocke \etal (1995\markcite{sto95}); b=Walter, Wolk \&
Neuh\"auser (1996\markcite{wwn96});
c= Neuh\"auser \etal (1997\markcite{n97}); d=Campana, Mereghetti \&
Sidoli (1997\markcite{cms97}); e=
Walter \& Matthews (1997\markcite{wm97}); f=Haberl \etal
(1997\markcite{ha97}); g=Motch \& Haberl (1998\markcite{mh98});
h = Kulkarni \& van Kerkwijk (1998\markcite{kvk98}); i= Schwope \etal
(1999)\markcite{sch99}; l = Motch
\etal (1999\markcite{mo99}); m = Haberl, Motch \& Pietsch
(1998\markcite{hmp98}); n = Haberl, Pietsch \& Motch 
(1999\markcite{hpm99})}
\enddata
\end{deluxetable}

Although present observations leave little doubt that these sources are
associated with isolated NSs, the exact nature of the mechanism powering
their emission is still controversial: either accretion of the ISM onto
old neutron stars or thermal radiation from middle--aged, cooling NSs. As
we discuss in section~\ref{avsc}, the emission properties of these two
classes of objects are rather similar and the present data cannot provide
convincing evidence in favor of either possibility.

\subsection{MS 0317.7-6647}\label{ms0317}

This source was the first to be proposed as an accreting ONS, and,
although its association with a neutron star was never supported
by further observations both in X--rays and in the optical, we
include it in our list.

MS 0317.7-6647 was already present in the Einstein Medium
Sensitivity Survey ( Gioia \etal 1990\markcite{gio90}) and then
re--observed with ROSAT by Stocke \etal (1995\markcite{sto95}). It
is a soft, very weak ($\sim 0.03$ count$\, \rm s^{-1}$) source in
the field of the nearby galaxy NGC 1313. From the three available X-ray images
there is evidence that this source is variable over the time scale of years.   
The only possible optical
counterpart has $m_V = 20.8$, which brings the X--ray to optical
flux ratio to $f_X/f_V\gtrsim 60$. The simultaneous lack of radio
emission seems to rule out the possibility of an extreme BL Lac
object leaving only a compact object as a reasonable option.
Identifying the source with an  X--ray binary in NGC 1313
leads to a luminosity $L\sim 10^{40} \ {\rm erg\, s^{-1}}$ and
requires a mass of the compact object $\sim 50\MS$. Such a large
value for the black hole mass, and the high Galactic latitude of
the source, which makes it unlikely to be an X--ray binary in our
own galaxy, were taken as clues in favor of the isolated NS
option. The PSPC spectrum, which is well fitted by a blackbody
with $T\sim 200$ eV, and  the derived luminosity, $L\sim 1.7\times
10^{30} \ {\rm erg/s}$ placing the object at $100$ pc, are fully
compatible with those of an isolated, nearby neutron star.
The  presence of a galactic IR cirrus cloud
in the field of view of the source
supports this possibility, while 
the sign of variability argues against the hypothesis of a cooling neutron
star (Brazier \& Johnston 1999\markcite{brj99}).

\subsection{RXJ 185635-3754}\label{1856}

RXJ 185635-3754 is a rather bright source, already present in  the
EINSTEIN slew survey, and observed with ROSAT by Walter, Wolk \&
Neuhauser (1996\markcite{wwn96}). The PSPC count rate, $\sim
3.6$ count$\, \rm{s}^{-1}$, corresponds to an X--ray  luminosity
$L \sim 5\times 10^{31} \ {\rm erg\, s^{-1}}$ for a distance of
100 pc. The observed spectrum is very soft and the best fit with a
blackbody gives a temperature of about 60 eV. The original claim
that this source is an isolated neutron star was supported by the
absence of an optical counterpart down to $m_V\sim 23$, giving
$f_X/f_V> 7000$.

Further X--ray and optical observations were performed by
Neuh\"user \etal (1997\markcite{n97}) and Campana, Mereghetti \& Sidoli
(1997\markcite{cms97})
in an attempt to identify the optical counterpart and to assess
its distance.  The source appears projected against the molecular cloud
R CrA and the column density derived from X--ray data ($N_H\sim
2\times 10^{20} \ {\rm cm}^{-2}$) is less than that in the cloud
itself. The cloud is estimated to be $\sim 120$ pc away, so the
source is likely to be closer. X--ray spectral fits with more
sophisticated models of cooling and accreting NS atmospheres were
presented by Pavlov \etal (1996\markcite{pztn96}) and Campana,
Mereghetti \& Sidoli (1997\markcite{cms97}). Unmagnetized cooling spectra
with different
chemical compositions predict an optical flux that is either too large or
too small, while both magnetized cooling and unmagnetized
accreting hydrogen models are consistent with the data only if the
source is as close as $\sim 10$ pc, and this seems difficult to
reconcile with the derived values of the column density.

The real nature of RXJ 185635-3754 became even more puzzling when
Walter \& Matthews (1997\markcite{wm97}) reported the discovery of
a possible optical counterpart with HST. Their candidate, selected
on the basis of positional coincidence and its blue colour, is
a faint, stellar--like object with $m_V\sim 25.6$, giving an
X--ray to optical flux ratio $\sim 75000$. They determined the
optical flux at two wavelengths, $\lambda = 3000, 6060$ A, and
both points lie above the Rayleigh--Jeans part of the blackbody
spectrum which best--fits the X--rays. The presence of a definite
optical excess, about a factor 3, points toward a more complex
spectrum and no definite explanation for this has been found up to
now (see section \ref{avsc} for details).

\subsection{RXJ 0720.4-3125 and RX J0420.0-5022}\label{0720}

RXJ 0720.4-3125 was observed for the first time with ROSAT by
Haberl \etal (1996\markcite{ha96}, 1997\markcite{ha97}). The
source was detected with the PSPC at a level  of $\sim 1.7$
count$\, \rm{s}^{-1}$ and the soft spectrum is well fitted by a
blackbody at $T\sim 80$ eV. The column density derived from the
fit is low, $N_H\sim
 10^{20} \ {\rm cm}^{-2}$, placing the source at a relatively close
distance, $d\approx 100$ pc, with a luminosity $L\sim 2.6\times
10^{31}(d/100 \, {\rm pc})^2 \ {\rm erg\, s^{-1}}$. Preliminary
optical searches produced a limiting magnitude for the counterpart
of $m_V> 21$, implying a flux ratio $> 500$.

Although the X--ray properties make RXJ 0720.4-3125 and RXJ
185635-3754 quite similar, RXJ 0720.4-3125 
has been found to pulsate in X--rays, with a
period  $P=8.39$ s. If indeed this source is powered by accretion,
the knowledge of the luminosity and of the period allows an
estimate of the magnetic field strength, as already pointed out by
Haberl \etal (1997\markcite{ha97}). The condition that the corotation
radius is
larger than the Alfv\`en radius (see eq. [\ref{psec}])
implies $B\lesssim 10^{10}$ G, indicating a substantial decay of
the magnetic field in this object.

Haberl \etal (1997\markcite{ha97}) suggested that such values of
$B$ and $P$ may be the result of common envelope evolution in a
high mass X--ray binary. In this case the history of RXJ
0720.4-3125 would bear resemblance to that of the Anomalous
X--ray Pulsars (AXPs; Mereghetti \& Stella 1995\markcite{ms95})  
if the last are interpreted as accreting from a residual circumstellar disk  
(van Paradijs, Taam \& van den
Heuvel 
1995\markcite{vp95}). Their observed periods,
in the 5--10 s range, are surprisingly similar to that of RXJ 0720.4-3125. 
Konenkov \& Popov
(1997\markcite{kp97}) and Wang (1997\markcite{w97}; see also Colpi
\etal 1998\markcite{elu98}) have shown that the present position
of RXJ 0720.4-3125 in the $B-P$ plane is consistent with the long
term evolution ($\sim 10^9-10^{10}$ yr) of a slow ($\sim 20$ \kms)
NS, born with canonical values of the magnetic field and period,
which experienced field decay over a timescale $\gtrsim 10^8$ yr.
The two evolutionary scenarios (common envelope or spontaneous
decay) will lead to the same kind of object, i.e. a slow,
accreting NS, but on quite different timescales, $\lesssim 10^7$
vs. $10^9-10^{10}$ yr.

An alternative explanation to the observed properties of RXJ
0720.4-3125 was recently proposed by Heyl \& Kulkarni
(1998\markcite{hk98}; see
also Wang 1997\markcite{w97} and Heyl \& Hernquist 1998\markcite{hh98}).
In this picture
the
source is powered by the release of internal energy, a magnetar
kept hot by magnetic field decay. For $B\sim 10^{14}$ G the time
required to cool down the NS to the observed temperature is $\sim
10^5$ yr, so contrary to the accretion scenario, RXJ 0720.4-3125
should contain a young NS with a decaying field. Heyl \& Kulkarni
argued that the source could be the descendant of an anomalous X-ray pulsar.
In this interpretation, AXPs are a
class of isolated neutron stars with unusually high magnetic
fields, powered by its decay (Thompson \& Duncan  1996\markcite{thdun96};
see also Colpi, Geppert, \& Page 1999\markcite{cgp99}). 

Recent follow--up optical observations (Motch \& Haberl
1998\markcite{mh98};
Kulkarni \& van Kerkwijk 1998\markcite{kvk98}) led to the identification
of a
possible optical counterpart, a faint, bluish object with $m_B\sim
26.6$. Again the identification is based on positional coincidence
and colour and gives  $f_X/f_V\sim 2\times 10^5$. The observed
optical flux, similarly to RXJ 185635-3754, exceeds that predicted
by the blackbody spectrum which fits the ROSAT data by a factor
$\sim 5$ (Kulkarni \& van Kerkwijk 1998\markcite{kvk98}).

RX J0420.0-5022 is the last isolated NS candidate discovered so far and the 
only one, together with RX J0720, to exhibit a modulation in the X--ray flux
with a period $\sim 22.7$ s (Haberl, Pietsch \& Motch 1999\markcite{hpm99}). 
It has the lowest observed count rate ($\sim 0.11$ PSPC count$\, 
\rm{s}^{-1}$), but shows all the distinctive features common to other 
sources in this class: a thermal spectrum with $T\sim 57$ eV, 
$N_H\sim 2\times 10^{20} \ {\rm cm}^{-2}$ which give $L\sim 2.7\times 10^{30}
(d/100 {\rm pc})^2\ {\rm erg\, s}^{-1}$. NTT optical observations revealed
no unusual object in the X--ray error box down to a magnitude $m_B\sim 25$, 
impliying an X--ray to optical flux ratio larger than $\sim 20000$. 

\subsection{RBS1223, RBS1556 (RX J1605.3+3249) and RX J0806.4-4132}
\label{rbs}

 RBS1223 and RBS1556 are two rather bright sources,
$\sim 0.3$ and $\sim 0.9$ PSPC count$\, \rm{s}^{-1}$ respectively,
detected in the ROSAT Bright Survey (RBS) and then observed in
follow--up pointings by Schwope \etal (1999\markcite{sch99}) and
Motch \etal (1999\markcite{mo99}). Both sources have similar
properties, a thermal X--ray spectrum with little absorption
($T\sim 100$ eV, $N_H\sim 10^{20} \ {\rm cm}^{-2}$). Optical
images  taken at Keck show no possible optical counterpart down to
$m_B\sim 26$ for RBS1223 (Schwope \etal 1999\markcite{sch99}) and
to $m_V\sim 24.2$ for RBS1556 (Motch \etal 1999\markcite{mo99}),
giving a lower limit for the X--ray to optical flux ratio in
excess of 10000.

RX J0806.4-4132 has been serendipitously discovered by Haberl,
Motch \& Pietsch (1998\markcite{mp98}) in archival PSPC observations of a
candidate supernova remnant. The quite large angular distance from
the source makes a physical association unlikely. The source has
all the distinctive features of the other candidates, the spectrum
is thermal with $T\sim 80 $ eV and the column density is low,
$N_H\sim 2.5\times 10^{20}  \ {\rm cm}^{-2}$. Preliminary optical
searches revealed no unusual object in the X--ray error box down to a
blue magnitude $\sim 24$.

\subsection{Results from ROSAT Surveys}\label{surveys}

Belloni, Zampieri \& Campana (1997\markcite{bzc97}) have performed
a systematic search for ONSs in two molecular clouds in Cygnus,
the Rift and OB7, by analyzing archive ROSAT PSPC pointings; this is the only
deep field survey available up to now.
Observations cover only a small fraction of the clouds, $\sim
5\%$, and contain 109 sources above a flux limit of $\sim 0.0015$
count$\, \rm{s}^{-1}$. For 105 of them an optical counterpart was
identified leaving 4 NOIDs, with no counterpart above $m_R\sim
20$. Although presently inferred values of $f_X/f_V\sim 1$ are not
large enough to qualify these sources as strong ONS candidates,
their positional coincidence with dense clouds together with their
hardness ratio $\gtrsim 0.3$ make them worthy of future
investigations. Belloni, Zampieri \& Campana estimated the number
of detectable ONSs in the searched area following the criteria of
section 5.1, and found that it exceeds
the number of NOIDs by a factor 3--10. Since ONSs can account only
for a fraction of NOIDs this results provides a clear indication
that theoretical figures produced so far overestimated the actual
number of detectable sources. A similar conclusion follows from
the comparison between the predicted number of resolved sources,
$\sim 10 \ {\rm deg}^{-2}$ and that of NOIDs in high--latitude
PSPC pointings, $\sim 30 \ {\rm deg}^{-2}$ discussed by Zane \etal 
(1995\markcite{ons}) in connection with the possible contribution of ONSs
to the diffuse X--ray background. As they noted, this would imply that
one NOID in three should be an ONS.

A more complete analysis of a larger sample region in Cygnus,
$\sim 64.5 \ {\rm deg}^2$, comprising a large part of the cloud
OB7, by Motch \etal (1997\markcite{mo97}) using RASS data lend
further support to this idea. The survey is complete to about 0.02
count$\, \rm s^{-1}$ and, at this flux level, 68 sources are
detected. Catalogue searches and optical follow--up observations
have shown that the vast majority of these sources are associated
with active coronae (F--K and M stars). There are 8 NOIDs and they
do not seem to be correlated with the denser regions of the cloud.
Again this figure is lower than the estimated number of ONSs
emitting above 0.02 counts/s ($\sim 10$ according to Zane \etal
1995\markcite{ons}).

A definite confirmation of the paucity of detectable ONSs in ROSAT
images came from the work of Danner (1998a,
b\markcite{d98a}\markcite{d98b}). He searched both high--latitude
molecular clouds and dark clouds in the Galactic plane for soft,
thermal X--ray sources using RASS data. The high--latitude cloud
sample used by Danner is that originally investigated by Magnani,
Blitz \& Mundy (1985\markcite{mbm85}) and the searched area is
$\approx 125  \ {\rm deg}^{2}$, substantially larger than that
occupied by the clouds themselves, $\sim 64.5 \ {\rm deg}^{2}$.
The total number of sources within the cloud boundaries is 89 and
the survey should be complete to 0.012 count$\, \rm{s}^{-1}$. Of
the 89 sources 54 are firmly identified with AGNs or stars, while
for those remaining one, or more, plausible counterparts are
reported. The claim is that at most one source could have escaped
the identification program, so the derived upper limit on the
projected density of ONSs is $\sim 0.016
 \ {\rm deg}^{-2}$. This figure is smaller than the number expected,
$\approx 0.15
 \ {\rm deg}^{-2}$ according to Zane \etal (1995\markcite{ons}), by
at least one order of magnitude.
In his dark cloud sample, which coincides with that analyzed by
Zane \etal (1995\markcite{ons}), only 16 sources met the selection
criteria and all of them have a firm, or at least plausible,
optical counterpart, in most cases a bright star. The only
exception is RX J1856.6-3754, but no other source with similar
properties was found. Since the dark cloud sample covers an area
of $\sim 1600 \ {\rm deg}^{2}$ and only one isolated NS candidate
is detected, the ONS projected density is $\lesssim 6\times
10^{-4}  \ {\rm deg}^{-2}$ at a count rate of 0.05 count$\,
\rm{s}^{-1}$. A direct scaling of the results in table 5 of Zane
\etal (1995\markcite{ons}) to a fixed (i.e. independent of the
cloud) threshold of 0.05 count$\, \rm{s}^{-1}$ gives 50 detectable
sources in the same area, nearly two orders of magnitude in excess
of the observed upper limit.

\section{Accreting versus Cooling Neutron Stars}\label{avsc}

According to the canonical scenario, soon after it is born in a type II 
Supernova event, a NS is very
hot ($T\sim 10^{11}$ K), has large magnetic field ($B\gtrsim
10^{12}$ G) and short period ($P\sim 0.01$ s). However, neutrino
emission quickly bring the temperature down
to $\sim 10^{6.5}-10^6$ K in about $10^3-10^4$ yr (see e.g. Page
1998\markcite{p98}). For the next $\sim 10^5$ yr, the star
temperature stays roughly constant (at $\lesssim 10^6$ K) and the NS
can be observed also in the X--ray band as a weak, soft, thermal
source. X--ray emission from radio pulsars and from the
$\gamma$--ray pulsar Geminga has been convincingly observed (see
e.g. \"Ogelman 1991, 1995\markcite{og91}\markcite{og95}; Foster
\etal 1996 \markcite{fo96}; Bignami \& Caraveo 1996\markcite{bc96}
for reviews; see also Becker \& Tr\"umper 1998\markcite{bt98}).
Many of these sources show a two--component spectrum and, while the origin
of the hard, non--thermal component remains uncertain, the soft
component almost certainly originates from the NS surface.
Fitting this thermal component with a blackbody gives $T_{eff}
\sim 10^5-10^6$ K, with lower temperatures associated with older
pulsars (see e.g. Pavlov \etal 1995\markcite{pszm95}).

Because of the typical values of $T_{eff}$, $L$ and of the thermal
spectrum, a middle--aged, radio--quiet, cooling object appears
quite similar to an accreting ONS in soft X--rays. This introduces
an ambiguity and the association of the isolated NS candidates
discovered so far with either an old or a (relatively) young
object remains a matter of lively debate. It is therefore
extremely urgent to improve our theoretical understanding of these
two classes of sources, looking, in particular, for spectral
signatures which can enable us to discriminate between them.

In principle, the lack of pulsations in the X--ray data could be
used as an argument against the presence of a high magnetic field.
This, in turn, may suggest that the field has decayed, favoring an
interpretation in terms of an old NS (see section \ref{paucity}).
Unfortunately, this is far from being conclusive since the
co--alignment between the rotation axis or the line of sight and
the magnetic axis would produce no pulsations even for high $B$.
On the other hand, field decay in isolated objects is highly
questionable, so the presence of pulsations is even less
conclusive in estimating the age of the NS.

At least one candidate, RX J1605.3+3249 for which X--ray measures
cover a substantial time interval, shows a remarkably constant
flux on various timescales (weeks to years). Motch \etal 
(1999\markcite{mo99}) argued that this can be taken as a hint in
favor of the cooling picture, because accretion instabilities
should lead to X--ray variability  on a free--free timescale which
is $\approx$ 1 yr under typical conditions. 
Also, ISM dishomogeneities on the smallest scales may give rise to
variations in the accretion rate. At a velocity of
$\sim 40$ \kms, an ONS would travel a distance $\sim 100$ AU in the
time separating the first and last ROSAT observation of  RX
J1605.3+3249 ($\sim 7.5$ yr). 

The (expected) similarity between accreting (low--luminosity) and
cooling NSs is not surprising. In both cases models are based on a
geometrically thin, static atmosphere in LTE which has roughly the
same thermodynamical properties. The only difference is the energy
input, either the heat released by the impinging ions or thermal
radiation emanating from the NS crust.

The problem of calculating the spectrum emerging from cooling NSs
has been widely investigated by a number of authors, both for
different chemical compositions and for low and high magnetic
fields (see e.g. Romani 1987\markcite{rom87}; Miller \&
Neuh\"auser 1991\markcite{mn91}; Shibanov \etal
1992\markcite{sh92}; Miller 1992\markcite{mil92}; Pavlov \etal
1994, 1995\markcite{pav94}\markcite{pszm95};
Zavlin \etal 1995\markcite{zav95}; Zavlin, Pavlov \& Shibanov
1996\markcite{zav96}; Rajagopal \& Romani 1996\markcite{rr96};
Pavlov \etal 1996\markcite{pztn96}; Rajagopal, Romani \& Miller
1997\markcite{rrm97}). These studies showed that emerging spectra
are not too different from a blackbody at the star effective
temperature. For $B\lesssim 10^{10}$ G, spectra from pure H
atmospheres exhibit a distinctive hard tail that becomes less
pronounced when the magnetic field is $\sim 10^{12}-10^{13}$ G.
However, as already mentioned in section 3.2, this feature cannot be
taken as characteristic of a
cooling object, since similar conclusions were reached for the
spectrum emitted by low--luminosity, low--field, accreting NSs
(Zampieri \etal 1995\markcite{ztzt95}; Zane, Turolla \& Treves
1999\markcite{mag}).

While a cooling atmosphere is likely to be rich in heavy elements
which results from the supernova explosion and the subsequent
envelope fallback, the assumption of a pure hydrogen composition,
although crude, is plausible for an accretion atmosphere. In this
case, in fact, incoming protons and spallation by energetic
particles in the magnetosphere may enrich the NS surface with
light elements (mainly H) and, owing to the rapid sedimentation,
these elements should dominate the photospheric layers (Bildsten,
Salpeter \& Wasserman 1992\markcite{bsw92}). As detailed
calculations have shown, the overall emission strongly depends on
the chemical composition. This led to the suggestion that the
comparison between observed and synthetic X--ray spectra may probe
the chemistry of the NS crust and, indirectly, shed light on the
powering mechanism.

Additional information on the nature of these sources may come
from optical observations. In fact, accretion spectra have been
found to exhibit a characteristic excess over the Rayleigh--Jeans
tail of the best--fitting (planckian) X--ray distribution, which
is not shared by cooling models with similar parameters (see Figure
2; Zane,
Turolla \& Treves 1999\markcite{mag}). The presence of an optical
bump is due to the fact that in the outermost layers the heating
produced by accretion cannot be balanced by free--free cooling.
The temperature then rises until $T\sim 10^7-10^8$ K, when Compton
cooling becomes efficient. The presence of this hot ``corona''
does not influence the propagation of X--ray photons, which come
from the inner layers, but makes the optical photosphere slightly
hotter than in the cooling case.

The observation of an optical excess has been reported in the
spectrum of RX J18563.5-3754 (see section \ref{1856}). Although
the optical identification of RX J0720.4-3125 still lacks a
definite confirmation, it is interesting to note that the
counterpart proposed by Kulkarni \& van Kerkwijk
(1998\markcite{kvk98}) also shows a similar excess.

The multiwavelength spectrum of RX J18563.5-375 has been the
subject of various investigations. Models of cooling atmospheres
based on different chemical compositions fail to predict the
correct optical fluxes (Pavlov \etal 1996\markcite{psc96}),
while models with two blackbody components or with a surface
temperature variation may fit the $f_{3000}$ flux. In particular,
the fact that a simple model with $T\propto$ (latitude)$^{0.25}$
and with magnetic and rotational axes co--aligned can match the
observed spectral energy distribution (SED), favored the
interpretation of RX J18563.5-3754 as a cooling NS (Walter \&
Matthews 1997\markcite{wm97}). Recent spectra from non--magnetic
atmospheres with Fe or Si--ash compositions (Walter \& An
1998\markcite{an98}) may also provide a fit of both the X--ray and
optical data, although they still need a definite confirmation.
The magnetized model by Zane, Turolla \& Treves
(1999)\markcite{mag} indeed predicts an optical excess of about
the right order indicating that, given the considerable latitude
of the unknown parameters, the full SED may be consistent with the
picture of an accreting NS. However, no detailed fit for RXJ
185635-3754 has been attempted yet. It should also be noted that
accretion spectra, irrespective of details on their shape, are
intrinsically  harder because of the reduced emitting area (see
eq. [\ref{teffmag}]). Assuming $f\sim 0.01$, which is conservative
if the field is high and equation (\ref{teffmag}) is taken
literally, the luminosity corresponding to a peak energy of $\sim
100$ eV is $10^{28}$--$10^{29}$ erg$\, \rm s^{-1}$s, more than one
order of magnitude lower than what follows from observations
placing the source at $\sim 100$ pc.

There has been a preliminary report of a proper motion measurement
for RXJ 185635-3754  by the MPE group (Tr\"umper, private
communication; Fred Walter's web page). If confirmed, the value of
$\sim 0.34$ arcsec/yr would imply a projected velocity of $\sim
160\, (d/100\, {\rm pc})$ \kms. Such a high velocity is definitely
incompatible with the observed flux, if the source is powered by
accretion and $d\sim 100$ pc. Nevertheless, the accretion option
could still be consistent with the data placing the source at a
closer distance. For $d\sim 30$ pc, a velocity $\sim 50$ \kmss would give
both the correct values of the proper motion and the
luminosity. Besides, a distance of the same order was derived by
Campana, Mereghetti \& Sidoli (1997\markcite{cms97}) from the fit
of ROSAT data with the unmagnetized synthetic spectrum of Zampieri
\etal (1995\markcite{ztzt}). Two major objections still remain.
First, the observed column density is too large to be consistent
with $d\sim 30$ pc, taking into account that the local ISM is
underdense. Second, the probability to find a NS so close {\it
and} with such a low velocity must be extremely small, also in the
light of the recent result by Popov \etal (1999\markcite{pop99}).
Here we just mention that, at variance with cooling atmospheres,
all accretion spectral models were computed so far under the
assumption of complete H ionization and do not include the
contribution of the atmospheric neutral atoms to the column
density. A rough estimate shows that a small fraction of neutrals
($\lesssim 10\%$) is anyway expected and this may produce the
required absorption ``in loco''. As far as the second point is
concerned, it is worth noting that the chance of finding a slow
ONS at 30 pc is not much slimmer than that of seeing a fast,
cooling NS at 100 pc. Taking a total number of NSs $\sim 10^9$,
the average ONS separation is $\sim 15$ pc, so the probability of
finding it at 30 pc is essentially 1. According to Popov \etal
(1999\markcite{pop99}) the fraction of low--velocity NSs is less
than $1\%$. This gives a total probability of $\lesssim 1\%$.
Cooling NSs are short--lived, $t_{cool}\sim 10^5-10^6$ yr (Page
1998\markcite{p98}) and their average separation is larger, $\sim
270$ pc. Now the probability of finding a cooling NS with $V\sim
240$ \kmss is essentially unity, but the chance of having it inside
100 pc is therefore $\sim (100/270)^3\sim 0.05$. The
conclusion is that {\em
both} events are quite rare.

The problem of distinguishing between accreting and cooling
isolated NSs on the basis of a statistical analysis has been
recently faced by Neuha\"user \& Tr\"umper (1999)\markcite{nt99}.
They have shown that the $\log N$--$\log S$ curve for the observed
candidates is closer to that of coolers, at least if ONSs are
taken as a slow population as in TC and BM, and suggested that
higher mean velocities may yield better agreement with observed
upper limits. This particular point will be addressed in section
\ref{fast}.

\section{Why Are So Few NSs Detected?}\label{paucity}

As discussed in section 7, the first analyses of ROSAT survey data
are mainly negative as far as ONSs are concerned, leading to a
problem of paucity of detections. The theoretical predictions of
the ``pre--ROSAT'' epoch appeared to be quite optimistic, forcing
theoreticians to reconsider critically their starting assumptions.
While the ONS spectral properties are rather well established,
earlier statistical estimates rely on a number of crucial
assumptions concerning the velocity distribution of pulsars at
birth and the long term evolution of the magnetic field.  Previous
studies (see section \ref{oldest}) were based on the hypothesis
that nearly all ONSs have a residual magnetic field $\sim 10^9$ G
and are presently in the accretor phase. However, accretion is
just one of the different phases that a NS can experience.
Normally born in the ejector stage, the NS enters later the
propeller, and the accretor phase if the conditions are favorable.
A time--dependent magnetic field may therefore change the duration
of these phases and alter the statistics of visible ONSs.
The evolution of the $B$--field in isolated NSs is still a very
controversial issue. Little evidence comes from the pulsar
statistics, and one should rely only on theoretical 
models (see e.g. Srinivasan 1997\markcite{sr97}). Theoretical results are
far from being 
univocal and predict either exponential/power--law field decay
(Ostriker \& Gunn 1969\markcite{og69}; Sang \& Chanmugam
1987\markcite{san87}; Urpin, Chanmugam \& Sang 1994\markcite{urp};
Miri 1996\markcite{miri96}; Urpin, \& Muslimov 1992\markcite{um92};
Urpin \& Konenkov 1997\markcite{uk97}) or
little or no decay at all
within the age of the Galaxy (Romani 1990\markcite{ro90};
Srinivasan \etal 1990\markcite{sri90}; Goldreich \& Reisenegger
1992\markcite{gr92}; see also Lamb 1992\markcite{la92} for a
review). Statistical analyses based on observations of isolated
radio pulsars are consistent with no field decay 
over the pulsar lifetime (Narayan \& Ostriker
1990\markcite{no90}; Sang \& Chanmugam 1990\markcite{sc90};
Bhattacharya \etal 1992\markcite{bha92}).
However, this does not preclude
the possibility of field decay over longer timescales.  Different
approaches to pulsar statistics led, independently, to the
conclusion that, if the magnetic field decays, then it probably
does so over a timescale $\sim 100$ Myr (Hartmann \etal
1997\markcite{hal97}).

A step forward in the statistical analysis is therefore to assume
different laws for the time evolution of the field and to combine
the dynamical evolution of the simulated population with its
magneto--rotational evolution. This has been computed for the first time
by Popov et al. (1999\markcite{pop99}) and is discussed in section
9.2. In section 9.1, we briefly summarize the role played by the
magnetic field in affecting the NS magneto--rotational evolution.

It should also be mentioned that the Bondi--Hoyle accretion rate
(eq. [\ref{mdot}]) may be just an upper limit. As discussed by
Blaes, Warren \& Madau (1995)\markcite{bwm95} the EUV--X radiation
released in the accretion process can heat the surrounding medium,
producing an increase of the sound speed which, in turn, tends to
decrease $\dot M$. This effect, known as {\em preheating}, may
severely reduce the accretion rate in ONSs.

\subsection{The role of the magnetic field}\label{decay}

The permanence of a NS in the ejector (or propeller) phase depends
on the magnetic field strength and is influenced by its time
evolution. Livio, Xu \& Frank (1998\markcite{liv98}) and Colpi \etal
(1998\markcite{elu98}) were the first to recognize that a NS can
straggle  in the ejector or propeller state for a time longer than
the present cosmic time mainly as a consequence of magnetic field
decay. This may cause the paucity of accreting ONSs. On a
theoretical basis, the decay occurs either because of ohmic
dissipation of crustal currents sustaining the field (Sang \&
Chanmugam 1987\markcite{san87}; Urpin \& Konenkov 1997\markcite{uk97}), or
because of the migration of
the proton fluxoids from the core to the crust due to continuous
spindown (Srinivasan \etal 1990\markcite{srini90}).

\begin{figure}
\epsfxsize=\hsize \centerline{{\epsfbox{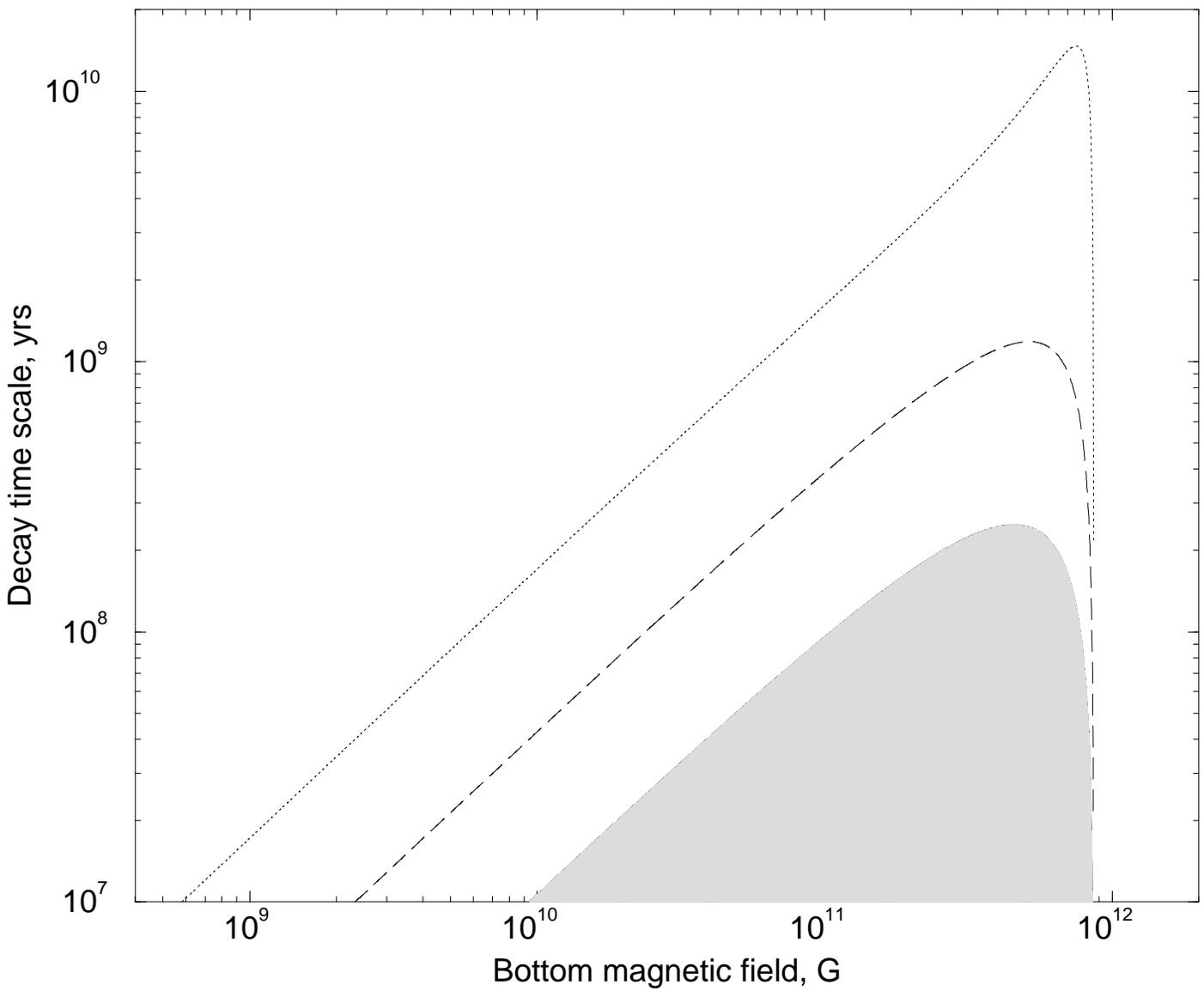}}}
\caption{The loci where the ejector lifetime equals the age of the
Galaxy (see text for details). \label{magnetic}}
\end{figure}

The idea that a NS may never become an accretor is rather simple
and can be illustrated by just estimating the duration of the
ejector stage $\tau_E$ and introducing a simple exponential decay law
for the field. Magnetic dipole braking in this phase keeps
the neutron star spinning down at a rate
\be
{\dot P}_{dipole}\sim 10^{-8}\left ({B(t)\over 10^{12} \rm{G}}\right )^2 P^{-1}{\rm
s\,\, yr^{-1}}
\ee
until  $P_{crit}$ is reached (see eq. [6]).
The critical period  is itself a function of time ($P_{crit}\propto B^{1/2}$)
and
decreases as $B(t)$ decays.
Two competing effects thus come into play: (i)  the spindown rate slows down
because of the weakening of the field, causing the NS to stay longer
in the ejector state;
(ii) $P_{crit}$ decreases with time and this acts in the opposite way,
making $\tau_E$ shorter.
If the field decays exponentially on a
scale $\tau_d$ from its initial value $B_0$ down to a minimum value
$B_b$  (the
presence of a relic field
is
mainly
suggested by the stability of the field in millisecond pulsars), the
duration of the ejector stage can be computed  analytically.
Following closely Popov \& Prokhorov (1999\markcite{poppro99}), the
ejector time scale
is
\be
\tau_E= \cases {-\tau_d\ln\left [
\displaystyle{{\tau_0\over \tau_d}}\left (1+
\displaystyle{{\tau_d^2\over \tau_0^2}}
\right )^{1/2}-1\right ] & $\tau_E<\tau_b$ \cr & \cr
\tau_b+ 
\displaystyle{{B_0\over B_b}} \left [\tau_0 
-{1\over 2}\tau_d
\left (1-e^{2\tau_b/\tau_d}\right ) \right ] & $\tau_E>\tau_b$, \cr}
\ee
where $\tau_{b}=\tau_d \ln(B_0/B_b)$ is the time at which the NS reaches
the minimum field and
\be
\tau_0\sim 4 n^{-1/2} v_{40}(B_0/10^{12}\rm {G})^{-1}\,\,\,\rm Gyr
\ee is the duration of the ejector phase for a constant field;
note that $\tau_0$ depends on $v$ and $n$. Figure (\ref{magnetic})
shows the loci where the time spent in the ejector phase equals
the age of the Galaxy, $\tau_E = 10$ Gyr, as a function of the
minimum field and of the decay timescale. Different curves refer
to initial fields of $5\times 10^{11}$, $10^{12}$, $2\times
10^{12}$ G and $v=40$ \kms, $n=1$ cm$^{-3}$ were assumed. The
``forbidden'' region, where $\tau_E > 10$ Gyr, lies below each
curve. As is apparent from figure \ref{magnetic}, the parameter
range for which the NS never leaves the ejector stage is
significant even for the rather low value of the star velocity we
adopted. It is interesting to note that unhalted decay (i.e.,
$B_b\to 0$)  would drive all NSs into the accretor phase, since
the Alfv\`en radius becomes smaller than the star radius as the
field decays.

Colpi \etal (1998\markcite{elu98}) considered a model in which the
spin evolution causes the core field to migrate to the crust where
dissipation processes drive ohmic field decay. They found that for
a characteristic time scale of $\sim 10^{8}$ yr (a value
consistent with the limits on the decay time scale inferred from
pulsar statistics) the number of ONSs that are at present in the
accretion stage can be reduced by a factor of 5 over previous
figures.  Most of the (low velocity) ONSs would now be in the
ejector or the propeller stages.

\subsection{Are ONSs Fast ?}\label{fast}
Pulsars have large kick velocities. Thus, if NSs at birth acquire
large mean speeds their distribution may be devoided of low
velocity stars. This may explain the paucity of observed ONSs. In
the light of new data on pulsars (see $\S$ \ref{distribuv}) a
re--analysis of the statistical properties of NSs is required.

Popov \etal (1999\markcite{pop99}) studied the effect of
an increasing initial mean velocity on the visibility of ONSs. In this 
investigation, the NS spin-down induced by dipole losses and their
interaction with the ambient medium (described using a realistic map of
the ISM in the Galaxy) has been followed, together with their dynamical
evolution in the Galactic potential, in unprecedented detail. Not
unexpectedly, the
ejectors were found to outnumber the accretors (and the
propellers): they comprise more than $95\%$ of all NSs at $\langle
V\rangle \sim 150$ \kms. The visible ONSs fall off rapidly when
$\langle V\rangle $ is above $200$ \kms, owing to the lack of low
velocity stars with increasing mean speed.

Using the present  upper limits on the number of detected ONSs
Popov \etal (1999\markcite{pop99}) were able to constrain the
pulsar velocity distribution independently of
radio observations. Taking a (generous)  upper limit of 10
accreting sources within 140 pc from the Sun, these authors
inferred a lower bound on the mean kick velocity $\langle
V\rangle$ of 200--300 \kms, corresponding to a dispersion
$\sigma_V\sim$ 125--190 \kms, in agreement with the findings of
Cordes \& Chernoff (1998\markcite{cc98}). In addition, these
results have been able to constrain the fraction of pulsars in the
low velocity tail which could have escaped pulsar statistics, to
$<0.05\%.$

\section{Conclusions}\label{conclu}

In the last decade old neutron stars accreting from the interstellar
medium have been actively sought, mainly in X--ray ROSAT images.
Although the quest produced half a dozen strong candidates, none of these
can be unambigously claimed as an accreting, isolated NS, so that the 
very existence of this class of sources is not observationally proven as yet.

The expected number of observed ONSs depend on the physical
properties of their progenitors, young NSs in the pulsar phase. The
key quantities are their velocity distribution and the law of magnetic
field decay, both subjects beeing frontline problems in pulsars
physics. 
Pulsars and ONSs would
be the A and $\Omega$ of the evolution, which brings the neutron star
through the intermediate phases of ejector and propeller. 

The physics of
propellers is poorly known and, especially in the case of very low
accretion rates, it is suspected that emission could
be non--stationary. If flaring episodes do occur, they may be of
critical importance in recognizing a population, which is
otherwise inaccessible because of its intrinsic low luminosity. 

Substantial improvements of our current understanding of the isolated NS
candidates discovered so far are expected from next generation X--ray 
satellites. Advanced spectroscopy with CHANDRA, XMM, and ASTRO E  will
provide 
high--resolution X--ray data which can reveal many properties of the 
NS atmosphere, like composition, gravitational redshift, temperature.
The comparison of high--quality spectra with models might indeed prove decisive
to assess the nature of these sources.
Optical observations with the largest telescopes, already operational, 
like KECK and VLT, would prove of the 
same importance in solving the riddle of isolated NS candidates. Without
the identification of optical counterparts they will remain X--ray bright
NOIDs. High--quality spectroscopy of the optical counterparts is needed 
to assess better the (possible) deviations from the Rayleigh--Jeans tail of 
the X--ray spectrum, like those reported in two candidates so far. 

Hopefully the next decade will bring us strong new candidates for isolated
NSs steadily accreting from the ISM, and the discovery of the previous
unsteady phase.

\acknowledgments

We would like to express our gratitude to a number of colleagues for 
many enlightening discussions. In particular we wish to thank 
Christian Motch, Fred Walter, Phil Charles, Sergio Campana, Sandro Mereghetti,
Sergei Popov, and Luca Zampieri for their helpful
suggestions on the
manuscripit. This work has been
partially supported by the European Commission under contract 
ERBFMRX-CT98-0195.

\clearpage

\clearpage

\clearpage

\end{document}